\documentclass[pre,aps,floatfix,superscriptaddress,two column]{revtex4-2}
\usepackage{graphicx}
\usepackage{dcolumn}
\usepackage{latexsym}
\usepackage{hyperref}
\usepackage{amsmath, amsthm, amssymb}
\usepackage{relsize}
\usepackage{epsfig}
\usepackage{bm}
\usepackage{geometry}
\usepackage{xcolor}
\usepackage{placeins}
\usepackage[normalem]{ulem}
\usepackage{diagbox}
\usepackage[normalem]{ulem}
\usepackage{tikz}
\usepackage{multirow}
\usepackage{makecell}

\usepackage{titlesec}

\titlespacing*{\section}
  {0pt}{*2}{*0.5}  % {left}{before}{after}

\titlespacing*{\subsection}
  {0pt}{*1.5}{*0.3}

\begin{document}

% \preprint{APS/123-QED}

%\title{Density dependent growth kinetics in two temperature mixture}
\title{Density-Induced Reentrant Coarsening in a Two-Temperature System}

\author{Partha Sarathi Mondal}
\email[]{parthasarathimondal.rs.phy21@itbhu.ac.in}
\affiliation{Department of Physics, Indian Institute of Technology (BHU) Varanasi, India 221005}

\author{Anish Kumar}
\affiliation{Department of Physics, Indian Institute of Technology (BHU) Varanasi, India 221005}

\author{Nayana Venkatareddy\textsuperscript{\textdagger}}
\affiliation{Department of Physics, Indian Institute of Science, C. V. Raman Ave, Bengaluru 560012, India}
\altaffiliation{Current address: Department of Physics, Technion- Israel Institute of Technology, Haifa 3200003, Israel.}

\author{Prabal K. Maiti}
\affiliation{Department of Physics, Indian Institute of Science, C. V. Raman Ave, Bengaluru 560012, India}

\author{Shradha Mishra}
\email[]{smishra.phy@iitbhu.ac.in}
\affiliation{Department of Physics, Indian Institute of Technology (BHU) Varanasi, India 221005}

\date{\today}% It is always \today, today,

\begin{abstract}
Understanding how nonequilibrium driving modifies phase-separation kinetics remains a fundamental challenge. Here we show that phase separation in a two-temperature system exhibits a striking density-induced reentrant coarsening behavior. Using Brownian dynamics simulations and a coarse-grained field-theoretic model, we find that the characteristic domain size grows as $L(t)\sim t^{1/z}$, displaying a reentrant sequence $(t^{1/3} \rightarrow t^{1/4}\rightarrow t^{1/3})$ with increasing density. While the low- and high-density regimes are governed by classical curvature-driven bulk diffusion, the intermediate-density regime exhibits anomalously slow growth. We show that this slowdown originates from a transport bottleneck arising from the interplay of particle diffusivity, particle availability, and attachment kinetics, which suppresses the effective mass flux between domains. Unlike equilibrium phase separation, where density primarily affects morphology and crossover scales, the two-temperature drive renders density a key control parameter for coarsening pathways. Our results uncover a nonequilibrium mechanism for anomalous domain growth in two-temperature systems.
\end{abstract}

\maketitle
% \section{Introduction}
Phase separation and domain coarsening are central to pattern formation across soft and biological matter, underpinning phenomena ranging from intracellular organization to driven colloidal assemblies \cite{bray1994theory,langer1980instabilities,berry2018physical}. In equilibrium systems, coarsening kinetics is commonly described in terms of different transport mechanisms, leading to well-characterized growth laws for the characteristic domain size \cite{bray1994theory,puri2009kinetics}. How such kinetic scaling is modified under nonequilibrium conditions - particularly when energy is injected locally - remains an area of active interest.\\
Active matter systems provide a natural framework to explore nonequilibrium coarsening \cite{wittkowski2014scalar,pattanayak2021ordering,pattanayak2021domain,das2018ordering,PhysRevLett.115.188302,yadav2025coarsening,paul2021clusters}, as they support phase separation mechanisms that need not rely on detailed balance. Among these, scalar active systems — where dissimilarities in the properties of constituent species lead to differences in their effective transport properties and drive activity — can be naturally described within a two-temperature framework and are relevant to a wide range of biological \cite{ganai2014chromosome,li2017double,xu2019direct} and synthetic mixtures \cite{bates1999block} and other natural systems \cite{triola2022model}.\\
Two-temperature induced phase separation (2-TIPS) in a binary mixture of `hot' and 'cold' particles has been studied extensively using both analytical methods \cite{grosberg2015nonequilibrium,grosberg2018dissipation,ilker2020phase,ilker2021long} and numerical simulations\cite{weber2016binary,mccarthy2024demixing, chari2019scalar, venkatareddy2023effect}.
While these studies established phase separation and diffusive coarsening in broad regions of parameter space, the role of density in controlling domain-growth kinetics remains largely unexplored. Since segregation and coarsening are both driven by interactions between the two species, density can fundamentally influence the transport processes responsible for domain growth. A central question, therefore, is whether density merely modifies crossover scales or can qualitatively reorganize the coarsening pathway itself.\\
In this Letter, we identify density as the key control parameter governing coarsening kinetics in two-temperature-induced phase separation (2-TIPS). By combining a coarse-grained field theory with Brownian dynamics simulations, we uncover a striking reentrant sequence of late time growth laws, $t^{1/3} \rightarrow t^{1/4} \rightarrow t^{1/3}$, revealing that the asymptotic coarsening dynamics are dictated primarily by density, while remaining largely insensitive to activity.\\
We show that the anomalous intermediate regime emerges from a restricted transport mechanism arising from the combined effects of the effective diffusivity of cold particles, their population in the dilute phase, and their attachment kinetics at the cluster interface. The same reentrant behavior is consistently reproduced by microscopic simulations and the coarse-grained theory, and is corroborated through scaling behavior of independent structural and interfacial measures. Beyond establishing a new route to controlling coarsening dynamics in scalar active systems, our work provides a unified physical framework linking microscopic transport processes to emergent growth laws in nonequilibrium phase separation.\\
% \section{Model}
To investigate how density influences domain-growth kinetics in 2-TIPS, we employ a coarse-grained (CG) description that explicitly resolves the densities of the cold and hot species, $\psi_c(\boldsymbol{r},t)$ and $\psi_h(\boldsymbol{r},t)$. Unlike conventional phase-separation models governed by a single conserved order parameter, the present framework retains the coupled dynamics of the two species and incorporates the nonequilibrium energy exchange arising from collisions between particles maintained at different temperatures. This exchange modifies the local effective temperature of each species through the density of the other species and is controlled by an activity parameter $\chi_c$, proportional to the temperature difference $(T_h-T_c)$.\\
The mean densities of the two species are chosen to be equal, $\langle\psi_c\rangle_{\boldsymbol{r}}=\langle\psi_h\rangle_{\boldsymbol{r}}=\psi_0$, where $\langle \cdots\rangle$ denotes averaging over all lattice points. Here, $\psi_0=0$ corresponds to the highest density in the system, while progressively smaller (negative) values of $\psi_0$ correspond to progressively lower densities.\\
To assess the robustness of the predictions of the coarse-grained model, we perform complementary microscopic (MI) Brownian dynamics simulations of a binary mixture of soft repulsive particles with diffusivity $D_{\rm hot}$ and $D_{\rm cold}$ $(D_{\rm hot}>D_{\rm cold}) \cite{weber2016binary}$. Non-equilibrium driving is controlled by the diffusivity ratio $(D=D_{\rm cold}/D_{\rm hot})$, which determines the relative temperature difference between the two species. We focus on symmetric mixtures with equal packing fractions $(\eta=\eta_c=\eta_h)$, so that the total density is varied through $(\eta_{\rm tot}=2\eta)$. Full details of the dynamical equations of coarse-grained (CG) as well as microscopic (MI) models and their numerical implementation are provided in the Supplemental Material (SM) Section I.

%%%%%%%%%%%%%-------growth exponent table---------------------------
%%%%%%%%%%%%%-------growth kinetics results for low density-------%%%%%%%%%%%
% \FloatBarrier
\begin{figure}[hbt]
    \centering
    \includegraphics[width=0.995\linewidth]{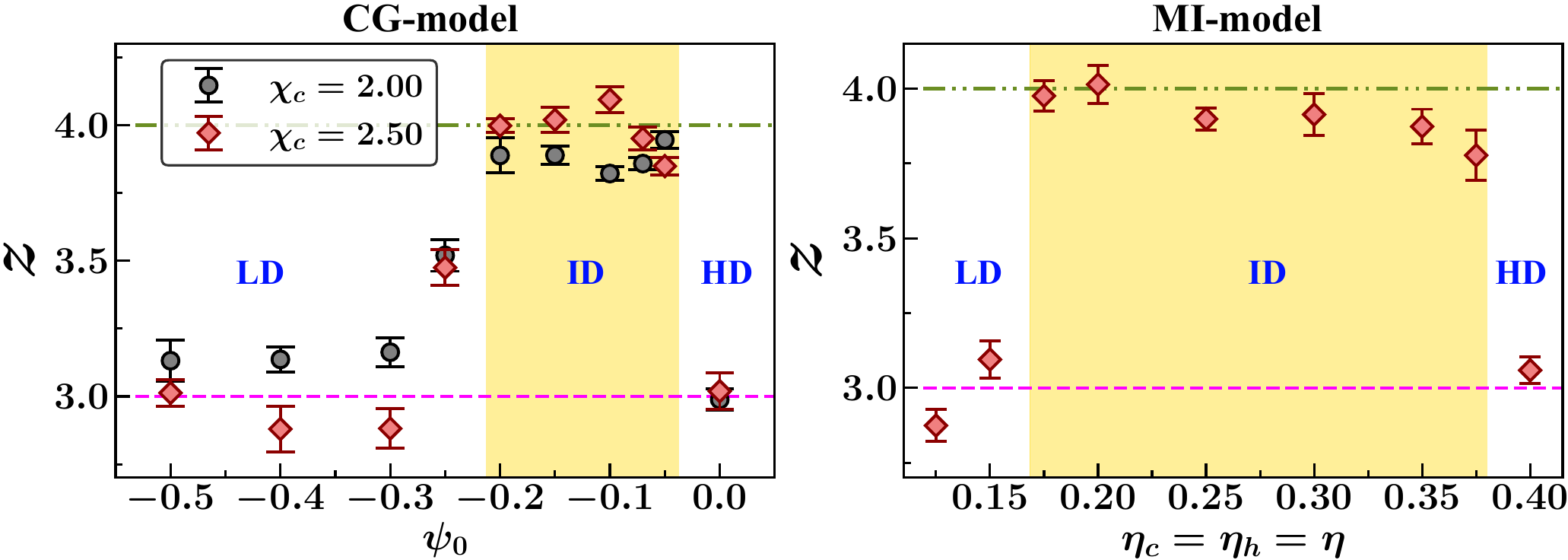}
    \caption{The variation of the late time growth exponent $z_{LT}$ from the CG-model in (a) and the MI-model in (b).}
    \label{fig:gexp_tab}
    % \vspace{-10pt}
\end{figure}

\begin{figure}[hbt]
    \centering
    \includegraphics[width=0.99\linewidth]{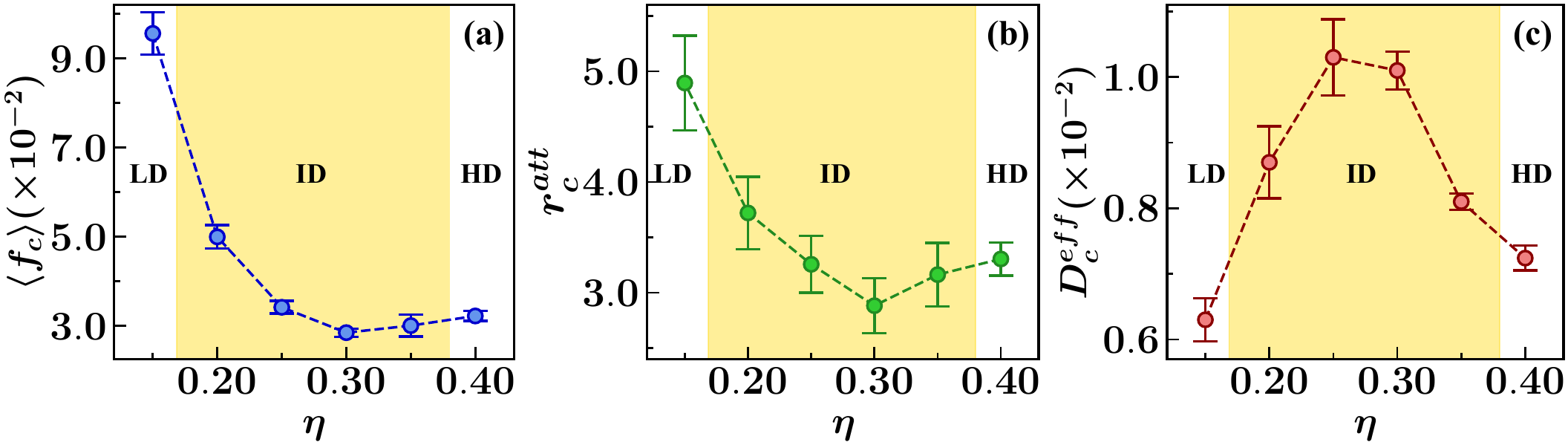}
    \caption{Average fraction of cold particles outside the clusters, $\langle f_c \rangle$ in (a), average attachment rate of cold particles from outside the clusters to the cluster interface, $r_c^{\mathrm{att}}$ in (b), and effective diffusivity $D_c^{\mathrm{eff}}$ in (c) as functions of the packing fraction $\eta$ obtained from microscopic simulations.}
    \label{fig:mech_mic}
\end{figure}
% \FloatBarrier
%%%%%%%%%%%%%%%%%%%%%%%%%%%%--------------------%%%%%%%%%%%%%%%%%%%%%%%%%%%%%%%%%%%%%%%%%%%%%%%%
%%%%%%%%%%%%%%%%%%%%%%%%%%%%%%%%%%%%%%%%%%%%%%%%%%%%%%%%%%%%%%%%%%%%

% %%%%%%%%%%%%%-------Length scale and scaling plots-------%%%%%%%%%%%
\begin{figure*}[hbt]
    \centering
    \includegraphics[width=0.99\linewidth]{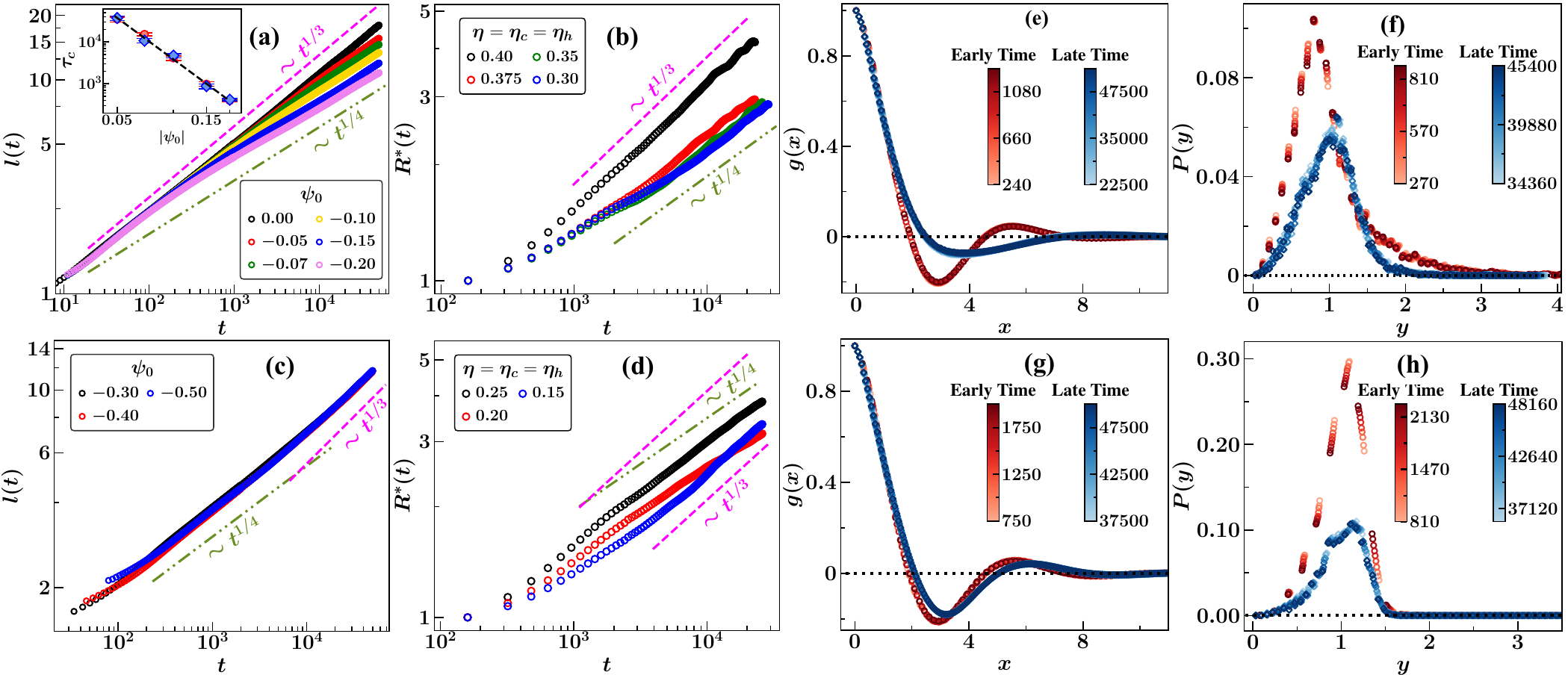}
    \caption{(a) $l(t)$ vs.\ $t$ from the CG-model for different off-criticalities $\psi_0$ at $\chi_{\mathrm{c}}=2.50$ in the HD $\to$ ID regime. (Inset) Crossover time $\tau_c$ vs.\ $|\psi_0|$; dashed line indicates the power-law fit $\tau_c \sim |\psi_0|^{3.30}$; (b) $R^*(t)$ vs.\ $t$ from the MI-model for different densities at $\chi_m=16.60$ in the HD $\to$ ID regime; (c) $l(t)$ vs.\ $t$ from the CG-model for various $\psi_0$ at $\chi_{\mathrm{c}}=2.00$ in the ID $\to$ LD regime; (d) $R^*(t)$ vs.\ $t$ from the MI-model for different densities at $\chi_m=16.60$ in the ID $\to$ LD regime. The standard deviation in the value of $l(t)$ and $R^{*}(t)$ obtained over independent realizations is of the size of the symbols in the respective cases. Scaled two-point correlation function $g(x)$ vs $x = r/l(t)$ [(e),(g)] and scaled cluster-size distribution $P(y)$ vs $y = R/\langle R \rangle$ [(f),(h)] obtained from the CG model. Panels (e),(f) correspond to $\psi_0 = -0.10$ and $\chi_c = 2.50$ (ID regime), while panels (g),(h) correspond to $\psi_0 = -0.50$ (LD regime) and $\chi_c = 2.00$. Parameters: system size $L_c=512$ for CG panels and no. of particles $N=15000$ for MI panels.}
    \label{fig:len_plot}
    % \vspace{-10pt}
\end{figure*}
% %%%%%%%%%%%%%%%%%%%%%%%%%%%%--------------------%%% %%%%%%%%%%%%%%%%%%%%%%%%%%%%%%%%%%%%%%%%%%%%%

\begin{figure}[hbt]
    \centering
    \includegraphics[width=0.99\linewidth]{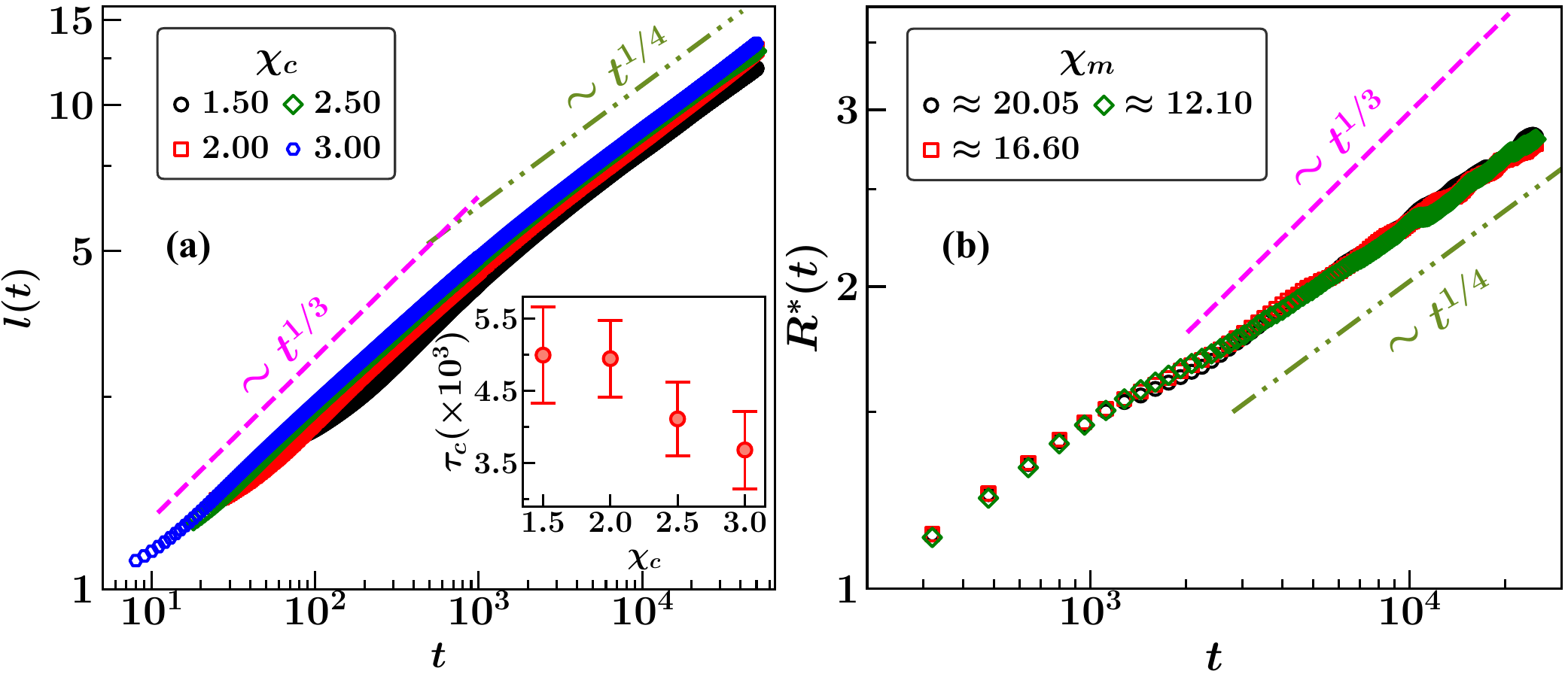}
    \caption{(a) Plot of $l(t)$ \emph{vs.} $t$ plot for different activity $\chi_c$ from the CG-model at $\psi_0 = -0.10$ in the ID regime. (inset) plot of crossover time $\tau_c$ \emph{vs.} activity $\chi_c$; (b) shows the $R^{*}(t)$ \emph{vs.} $t$ for different activity $\chi_m$ from the MI-model at $\eta = 0.30$ in the ID regime. The rest of the parameters for the CG and MI models are the same as in Fig.\ref{fig:len_plot}.}
    \label{fig:act_dep}
    % \vspace{-10pt}
\end{figure}

\begin{figure}[hbt]
    \centering
    \includegraphics[width=0.999\linewidth]{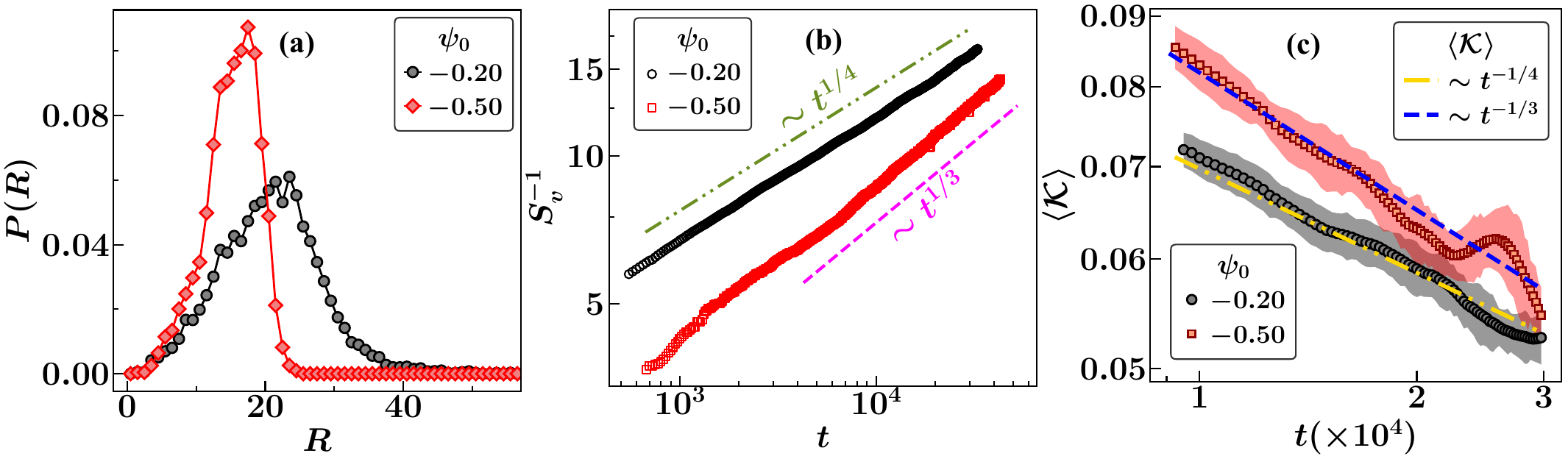}
    \caption{Comparison of morphological characteristics from the CG-model for different off-criticality $\psi_0 = -0.20$, and $-0.50$ for activity $\chi = 2.00$. (a) shows the plot of cluster size distribution $P(R)$ \emph{vs.} $R$. (b) shows time evolution of $S_{v}^{-1}$. Panel (c) shows the time evolution of mean curvature $\langle \mathcal{K} \rangle$. The rest of the parameters for the CG and MI models are the same as in Fig.\ref{fig:len_plot}.}
    \label{fig:interface_morphology}
\end{figure}

% \section{Results}
Starting from homogeneous initial conditions, the system undergoes 2-TIPS and forms cold-rich domains embedded in a hot-rich background (Fig.S1 in SM). The resulting morphologies of cold particle clusters depend strongly on density. At high densities, the domains are interconnected and highly convoluted \cite{venkatareddy2025growth}. At intermediate densities, the morphology consists of convoluted yet isolated domains. Upon further reducing the density, isolated droplet-like domains become the dominant morphological feature of the system. Representative snapshots illustrating these distinct morphologies are shown in Fig.S3 in the SM. The phase separation between the cold and hot particles is discussed in Sec.II in SM.\\
As time proceeds, the domains grow in size (Fig.S1 $\&$ MOV-I, II in SM). To quantify the temporal evolution of domains, we calculate the equal-time two-point correlation function $C(r,t)$ and structure factor $S(\mathrm{k},t)$ (Sec.IV in SM) of the phase-separation order parameter $\phi(\boldsymbol{r}, t) = \psi_{\mathrm{h}}(\boldsymbol{r}, t) - \psi_{\mathrm{c}}(\boldsymbol{r}, t)$. The average domain size is quantified by the characteristic length scale extracted from $S(\mathrm{k},t)$ in the CG model, denoted by $l(t)$, and by the radius of gyration of cold-particle clusters in the microscopic model, denoted by $R^{*}(t)$. The coarsening dynamics are then characterized by the temporal evolution of these length scales. The cluster detection algorithm for the MI-model is described in Sec.III in SM.\\
The temporal evolution of $L(t)$ reveals a remarkable density dependence of the coarsening kinetics, where $L(t)=l(t)$ for the CG model and $L(t)=R_g^{*}(t)$ for the MI model. In the scaling regime, the characteristic length grows algebraically as $L(t) \sim t^{1/z}$, where $z$ denotes the growth exponent. Figure \ref{fig:gexp_tab}(a-b) summarizes the variation of the asymptotic growth exponent $z$ with density for both the coarse-grained and microscopic models. Remarkably, $z$ exhibits a pronounced reentrant behavior. At high densities, domain growth follows the classical Lifshitz-Slyozov law \cite{bray1994theory}, $l(t)\sim t^{1/3}$\cite{venkatareddy2025growth}. Upon decreasing density, the growth slows down and crosses over to a $t^{1/4}$ regime over an extended intermediate-density window. A further reduction of density restores the conventional $t^{1/3}$ growth law. The excellent agreement between the coarse-grained and microscopic descriptions demonstrates that this reentrant behavior is robust. To systematically analyze this reentrant behavior, we divide the entire density range into three regimes (i) high density (HD) $\eta \gtrsim 0.40$ $\&$ $\psi_0 =0$, (ii) intermediate density (ID) $0.35 \le \eta \leq 0.20$ $\&$ $-0.20 \le \psi_0 \le -0.05$, and (iii) low density (LD) $\eta < 0.15$ $\&$ $-0.50 \le \psi_0 \le -0.30$. The different regimes are marked in Fig.\ref{fig:gexp_tab}(a-b) and Fig.\ref{fig:mech_mic}(a-c).\\
The non-monotonic variation of the growth exponent suggests a change in the transport mechanism controlling domain growth. The coarsening of domains is controlled by particle transport through the dilute phase outside the clusters, which in the present system is governed by the dynamics of the cold particles outside the cluster. The growth kinetics is set by the slower of the two processes, (i) bulk diffusion and (ii) interfacial attachment. To identify the mechanisms governing the coarsening kinetics, we calculate the attachment rate of cold particles from the dilute phase to the cluster surface, $r_c^{\mathrm{att}}$, the fraction of cold particles outside the clusters, $\langle f_c \rangle$, and the effective diffusivity of cold particles, $D_c^{\mathrm{eff}}$, which are straightforward to compute in the particle-based model. As shown in Fig.\ref{fig:mech_mic}, all three quantities show non-monotonic behavior with respect to packing fraction $\eta$. $\langle f_c \rangle$ is largest at low densities and decreases with increasing $\eta$, exhibiting a minimum in the intermediate density regime. Simultaneously, attachment rate $r_c^{att}$ shows a similar minimum at intermediate densities. In contrast, $D_c^{eff}$ exhibits a maximum in the intermediate density regime.\\
These observations indicate that, in the low-density regime, a large population of cold particles remains in the dilute phase, and the coarsening is governed by diffusion-limited transport of particles to the clusters, leading to the observed $~ t^{1/3}$ domain growth. \\
But on the other hand, for the intermediate densities, $D_c^{eff}$ is larger, hence growth is no longer diffusion-limited, but controlled by the attachment rate of cold particles at the surface of a cluster from the sea of the hot particles.  
In the intermediate density regime, although the cold particles in the dilute phase are more mobile, both the population of particles in the dilute region and the attachment rate to the cluster surface are reduced. Consequently, the growth is dominated by the surface attachment of particles to the clusters \cite{weber2016binary}, resulting in slower growth consistent with $\sim t^{1/4}$. At high densities, the growth again becomes diffusion-limited. In this regime, we also observe that cluster coalescence becomes important (MOV-3 in SM), and the combined effect of diffusive transport and coalescence leads to the recovery of $\sim t^{1/3}$ growth. \\
We now revisit the temporal evolution of the characteristic length scale and examine its behavior throughout the coarsening process in both the CG and MI models (Fig.\ref{fig:len_plot}(a-d)). In the HD regime, we observe $L(t) \sim t^{1/3}$, in agreement with our earlier results \cite{venkatareddy2025growth}. Upon decreasing density, coarsening at late times slows down, and in the ID regimes, both MI and CG models exhibit $L(t) \sim t^{1/4}$ growth at late times. In the CG-model, $l(t)$ \emph{vs.} $t$ plot exhibits two distinct scaling regimes: $t^{1/3}$ growth at early times to a slower $t^{1/4}$ growth at late times. Extensive finite-size analysis, spanning system sizes differing by $\mathcal{O}(10)$, confirms that the observed slower growth at late times is intrinsic and not a finite-size artifact (Fig.S7 in Sec.V in SM). The crossover is quantified by the temporal evolution of the effective growth exponent $\frac{1}{z_{eff}}=\frac{d\{\ln[l(t)] \}}{d\{\ln (t) \}}$ (Fig.S8 in Sec.VI in SM), from which the cross-over time $\tau_c$ is calculated. We find that $\tau_c$ decreases monotonically with decreasing density (increasing off-criticality), indicating that the crossover to the slower growth regime occurs progressively earlier within the ID regime. In contrast, no comparable crossover is observed in the MI model.\\ 
On further reduction of density, late time $t^{1/3}$ growth is recovered in the LD regime. In this regime, both the MI and CG models display two stage coarsening: an initial slower $t^{1/4}$ growth followed by a faster $t^{1/3}$ growth at late times. Finite-size analysis confirms that this crossover is intrinsic and not a finite-size artifact (Fig.S7 SM).\\
Further, we focus on the scaling characteristics to distinguish the $1/3$ and $1/4$ growth laws for ID and LD regimes, only using the coarse-grained model. In standard coarsening scenarios, a single power-law growth of the characteristic length scale is typically accompanied by dynamical scaling, reflected in time-invariant scaling forms of correlation functions and cluster statistics \cite{bray1994theory,toral1992droplet}.  Having established that distinct transport mechanisms operating across different density regimes give rise to a reentrant variation of the asymptotic growth exponent, we next examine whether the corresponding scaling behavior is also manifested in independent structural and interfacial observables. To this end, we first examine the scaling properties of the correlation function $C(r)$ and the cluster-size distribution $P(R)$. The details of cluster detection in the CG model are described in Sec.~III of the SM.\\
In the ID-regime, both the scaled correlation function $g(x)$, with $x=r/l(t)$, and the scaled cluster size distribution $P(y)$, with $y=R/\langle R\rangle$, exhibit qualitatively different scaling forms at early and late times, providing clear evidence for the presence of two distinct dynamical scaling regimes as shown in Fig.\ref{fig:len_plot}(e-f). Consistently, optimal data collapse of $C(r,t)$ is obtained upon rescaling distances as $r' = r\,t^{-1/\delta}$, with $\delta=3$ at early times and $\delta=4$ at late times (Sec.VII in SM). In the LD regime, a similar two-stage scaling behavior is observed for $g(x)$ and $P(y)$ as shown in Fig.\ref{fig:len_plot}(g-h). Moreover, the correlation functions exhibit optimal collapse with $\delta=4$ at early times, followed by a crossover to $\delta=3$ at late times (Sec.VII in SM).\\
The growth exponents and crossover behavior are largely insensitive to activity in both the CG and MI models, Fig.  \ref{fig:act_dep}(a-b). These results demonstrate that activity has a negligible impact on the coarsening exponents, despite playing a significant role in the emergence and maintenance of phase separation (Fig.S2 in SM).\\
The distinct growth regimes with the underlying structural evolution in the late time regime are also correlated to cluster statistics and interfacial measures. Fig.\ref{fig:interface_morphology} compares the cluster statistics and the interfacial measures for two representative off-criticalities $\psi_0 = -0.20$ (ID-regime) and $-0.50$ (LD-regime) from the CG-model.\\
Fig.\ref{fig:interface_morphology}(a) presents the unscaled CSD $P(R)$ in the late time regimes. For $\psi_0 = -0.20$, the distribution remains broad, indicating the presence of clusters spanning a wide range of sizes. In contrast, for $\psi_0 = -0.50$, $P(R)$ is narrower and more strongly peaked, implying a more uniform domain population. These differences reflect distinct patterns of structural organization in the two regimes.\\
To further probe the distinct asymptotic growth regimes, we calculate the characteristics interfacial length scale $S_v^{-1}$, defined as the inverse of interfacial length density in two dimensions\cite{marsh1996overview}, and the mean interfacial curvature $\langle \mathcal{K} \rangle$. The details of interface detection and curvature calculation are discussed in Sec.VIII in SM. As shown in Fig.\ref{fig:interface_morphology}(b), $S_v^{-1}$ grows as a power law with time, with exponent $\approx \frac{1}{4}$ for $\psi_0 = -0.20$ and $\approx \frac{1}{3}$ for $\psi_0 = -0.50$, respectively. Proportionately, the $\langle \mathcal{K} \rangle$ decays as $t^{-1/4}$ and $t^{-1/3}$ for the two cases, respectively [Fig.\ref{fig:interface_morphology}(c)]. Since the mean interfacial curvature provides a measure of typical domain size, $\langle l_{d} \rangle \propto \langle \mathcal{K} \rangle^{-1}$, the interfacial evolution corroborates the growth exponents obtained from the characteristic length scale. Together with the scaling behavior of the correlation function and the cluster-size distribution, these interfacial measures provide compelling evidence for the existence of distinct asymptotic growth exponents across different density regimes.\\
Further support for the argument is obtained from the coarse grained model, which also exhibits domain coarsening through the Ostwald ripening mechanism, similar to the microscopic model, wherein larger droplets grow at the expense of smaller ones. In the intermediate- and low-density regimes, the rate of shrinking of smaller droplets of radius $R$ follows the scaling $\frac{dR}{dt} \sim \frac{1}{R^3}$ and $\sim \frac{1}{R^2}$, respectively (Fig.S11 in Sec.IX in SM).\\
In TABLE.S1 (in Sec.X in SM), we summarize the scaling behavior of the key observables across different density regimes in the CG model.

%%%%%%%%%%%%%-------groth kinetics for different activity-------%%%%%%%%%%%
% \FloatBarrier
% \FloatBarrier
%%%%%%%%%%%%%%%%%%%%%%%%%%%%--------------------%%%%%%%%%%%%%%%%%%%%%%%%%%%%%%%%%%%%%%%%%%%%%%%%

% %%%%%%%%%%%%%-------statistical and interfacial characteristics-------%%%%%%%%%%%
%%%%%%%%%%%%%%%%%%%%%%%%%%%%%%%%%%%%%%%%%%%%%%%%%%%%%%%%%%%%%%%%%%%%%%%%%%%%%%%%%%%%

To summarize, using complementary coarse-grained and microscopic descriptions, we show that density acts as a control parameter for coarsening universality in 2-TIPS. The system exhibits a reentrant sequence of growth laws, with the characteristic domain size growing as $L(t)\sim t^{1/3}, t^{1/4}$, and $t^{1/3}$ as the density is varied from high to intermediate and then low values. The agreement between the two approaches demonstrates that this behavior is robust and independent of the microscopic implementation of the nonequilibrium drive. The distinct growth regimes are consistently reflected in the scaling properties of correlation functions, cluster statistics, and interfacial geometry, while the growth exponents remain largely insensitive to activity.\\
The observed reentrant kinetics originate from a density-driven change in the dominant transport mechanism governing domain growth. At low densities, coarsening is controlled by diffusion-mediated mass transport through the dilute phase, yielding the classical $t^{1/3}$ growth law. At intermediate densities, the reduced availability of cold particles and suppressed attachment kinetics at cluster interfaces generate a transport bottleneck, leading to the slower $t^{1/4}$ growth. Our results, therefore, identify density as a nonequilibrium control parameter that selects the transport mechanism and, consequently, the coarsening universality class.\\
The present framework can be extended naturally to include the inertial effect. In the current work, we have considered an overdamped system. Recent work has shown that, the presence of inertial effects can substantially accelerate coarsening in the low-density regime by introducing additional transport pathways \cite{venkatareddy2023phase}. Moreover, while the present study considers a two-temperature mixture in which the constituent species differ only in temperature, extending the framework to mixtures with additional asymmetries, such as particle shape or size, would elucidate how these asymmetries influence the kinetics of phase separation. 

\section*{Acknowledgement}
The authors thank Chandan Dasgupta for useful discussion during initial stage of the work. P.S.M. thanks UGC for the research fellowship. A.K. thanks PMRF, INDIA, for the research fellowship. S.M. thanks DST, SERB (INDIA), Project No.: CRG/2021/006945, MTR/2021/000438, and ANRF grant numbered ANRF/ARG/2025/008220/PS  for financial support. PKM thanks ANRF, India, for support through the JC Bose Grant (ANRF/JBG/2025/000227/PS).

\newpage

% \noindent
% \emph{Mechanism of domain growth}. 

% {\section{Discussion}\label{sec:dis}}

% \section{Acknowledgment}

\bibliography{references}% Produces the bibliography via BibTeX.

\end{document}